\documentclass[aps,prl,showpacs,floatfix,twocolumn,superscriptaddress,longbibliography]{revtex4-1}

\usepackage{color}
\usepackage{hyperref}
\usepackage{todonotes}
\usepackage{verbatim}
\usepackage{soul}
\usepackage{glossaries}
\usepackage{sidecap}

\def\mathbi#1{\ensuremath{\textbf{\em #1}}}
\def\Q{\mathbi{Q}}
\def\QCDW{\ensuremath{\mathbi{Q}_{\text{CDW}}}}
\def\QCDW{\ensuremath{\mathbi{Q}_{\text{CDW}}}}

\newacronym{BS}{BS}{bond-stretching}
\newacronym{RIXS}{RIXS}{resonant inelastic x-ray scattering}
\newacronym{LSCO}{LSCO}{La$_{2-x}$Sr$_{x}$CuO$_{4}$}
\newacronym{EPC}{EPC}{electron-phonon coupling}
\newacronym{CDW}{CDW}{charge-density wave}
\newacronym{SDW}{SDW}{spin-density wave}
\newacronym{FWHM}{FWHM}{full-width at half-maximum}
\newacronym{INS}{INS}{inelastic neutron scattering}

\begin{document}

\title{Nature of the charge-density wave excitations in cuprates} 

\author{J. Q. lin}\email[]{jiaqilin@bnl.gov}
\affiliation{Condensed Matter Physics and Materials Science Department, Brookhaven National Laboratory, Upton, New York 11973, USA}
\affiliation{School of Physical Science and Technology, ShanghaiTech University, Shanghai 201210, China}
\affiliation{Institute of Physics, Chinese Academy of Sciences, Beijing 100190, China
}
\affiliation{University of Chinese Academy of Sciences, Beijing 100049, China}
\author{H. Miao}
\altaffiliation[Present address: ]{Materials Science and Technology Division, Oak Ridge National Laboratory, Oak Ridge, Tennessee 37831, USA}
\affiliation{Condensed Matter Physics and Materials Science Department, Brookhaven National Laboratory, Upton, New York 11973, USA}
\author{D. G. Mazzone}
\author{G. D. Gu}
\affiliation{Condensed Matter Physics and Materials Science Department, Brookhaven National Laboratory, Upton, New York 11973, USA}

\author{A. Nag}
\author{A. C. Walters}
\author{M. Garc\'{i}a-Fern\'{a}ndez}
\affiliation{Diamond Light Source, Harwell Science and Innovation Campus, Didcot, Oxfordshire OX11 0DE, United Kingdom}

\author{A. Barbour}
\author{J. Pelliciari}
\author{I. Jarrige}
\affiliation{National Synchrotron Light Source II, Brookhaven National Laboratory, Upton, NY 11973, USA}

\author{M. Oda}
\author{K. Kurosawa}
\affiliation{Department of Physics, Hokkaido University, Sapporo 060-0810, Japan}
\author{N. Momono}
\affiliation{Department of Sciences and Informatics, Muroran Institute of Technology, Muroran 050-8585, Japan}

\author{K. Zhou}
\affiliation{Diamond Light Source, Harwell Science and Innovation Campus, Didcot, Oxfordshire OX11 0DE, United Kingdom}
\author{V. Bisogni}
\affiliation{National Synchrotron Light Source II, Brookhaven National Laboratory, Upton, NY 11973, USA}
\author{X. Liu}\email[]{liuxr@shanghaitech.edu.cn}
\affiliation{School of Physical Science and Technology, ShanghaiTech University, Shanghai 201210, China}
\author{M. P. M. Dean}\email[]{mdean@bnl.gov}
\affiliation{Condensed Matter Physics and Materials Science Department, Brookhaven National Laboratory, Upton, New York 11973, USA}

\date{\today}

\begin{abstract}
The discovery of \gls*{CDW}-related effects in the \gls*{RIXS} spectra of cuprates holds the tantalizing promise of clarifying the interactions that stabilize the electronic order. Here, we report a comprehensive \gls*{RIXS} study of \gls*{LSCO} finding that \gls*{CDW} effects persist up to a remarkably high doping level of $x=0.21$ before disappearing at $x=0.25$. The inelastic excitation spectra remain essentially unchanged with doping despite crossing a topological transition in the Fermi surface. This indicates that the spectra contain little or no direct coupling to electronic excitations near the Fermi surface, rather they are dominated by the resonant cross-section for phonons and CDW-induced phonon-softening. We interpret our results in terms of a \gls*{CDW} that is generated by strong correlations and a phonon response that is driven by the \gls*{CDW}-induced modification of the lattice.
\end{abstract}

\maketitle

\glsresetall 

\begin{figure}
\center
\includegraphics[width = 0.5\textwidth]{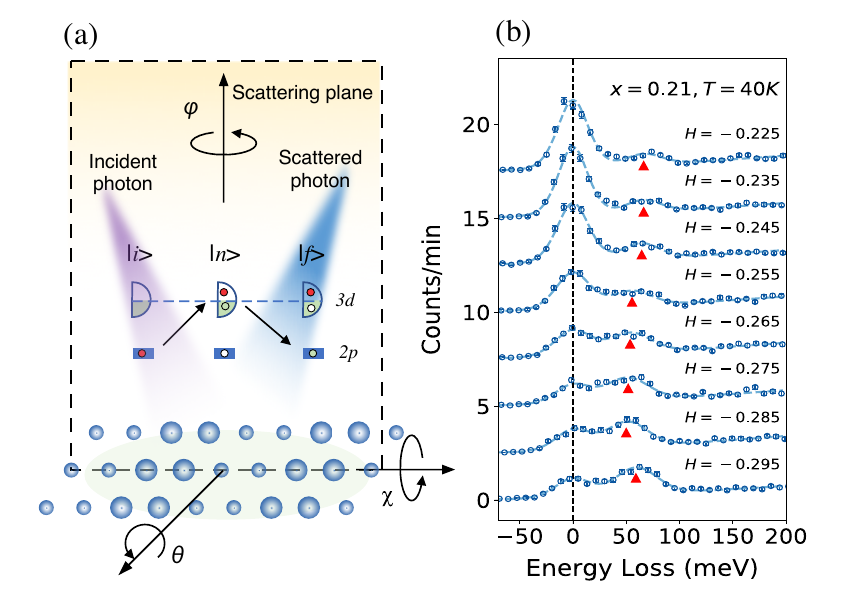}
\caption{\gls*{RIXS} process and typical low-energy \gls*{RIXS} spectra. (a) Schematic of the Cu \textit{L}-edge \gls*{RIXS} process and experimental setup. $|i\rangle$, $|n\rangle$ and $|f\rangle$ represent initial, intermediate and final states, respectively and solid (empty) circles represent occupied (unoccupied) states. $\theta$, $\chi$, and $\phi$ denote the sample rotations. (b) High-resolution ($\Delta E = 30$~meV) \gls*{RIXS} data for LSCO21 at $T=40$~K (blue circles) illustrating the principle components of the spectra: Quasi-elastic scattering (marked by the vertical dotted line) and a low-energy phonon excitation (indicated by red triangles). The fit, as described in the main text, is represented by the dashed blue line.}
\label{intro}
\end{figure}

Thirty years after the discovery of high-temperature superconductivity in the cuprates, there is still no consensus regarding the minimal set of interactions needed to describe the ``normal state'' from which superconductivity emerges \cite{Keimer2015}. Popular Hubbard and `$t-J$' theoretical models suggest that superconducting, \gls*{CDW}, and \gls*{SDW} states have similar ground state energies \cite{Zaanen1989, Poilblanc1989, Emery1990, Castellani1995, Corboz2014, Huang2017, Zheng2017}, which is consistent with experiments that reveal widespread interplay between all three states \cite{Tranquada1995, Ghiringhelli2012, Blanco-Canosa2014, Tabis2017, Comin2014, Croft2014, Thampy2014}. This complexity motivates experimental efforts to measure collective excitations associated with \gls*{CDW} order that should clarify the pertinent interactions. Resonant inelastic x-ray scattering (RIXS)\glsunset{RIXS}, as illustrated in Fig.~\ref{intro}(a), has an enhanced sensitivity to valence charge and phonon excitations \cite{Ament2011}. Several experiments indeed reported anomalies in cuprate \gls*{RIXS} spectra at the \gls*{CDW} wavevector (\QCDW{}), opening new routes to understand cuprate \gls*{CDW}s \cite{Dean2013, Miao2017, Chaix2017, Arpaia2019, Yu2019, Peng2019enhanced}. Uniquely isolating and interpreting \gls*{CDW}-effects in \gls*{RIXS} is, however, complicated as \gls*{CDW}s inevitably modify their host crystal lattice and thus the phonons. Compounding this problem, \gls*{RIXS} spectra near \QCDW{} have been conceptualized in several different ways including charge excitations \cite{Miao2019a, Arpaia2019, Yu2019}, momentum-dependent \gls*{EPC} \cite{Peng2019enhanced} and Fano-effects \cite{Chaix2017}.  

\begin{figure*}
\includegraphics[width = 1\textwidth]{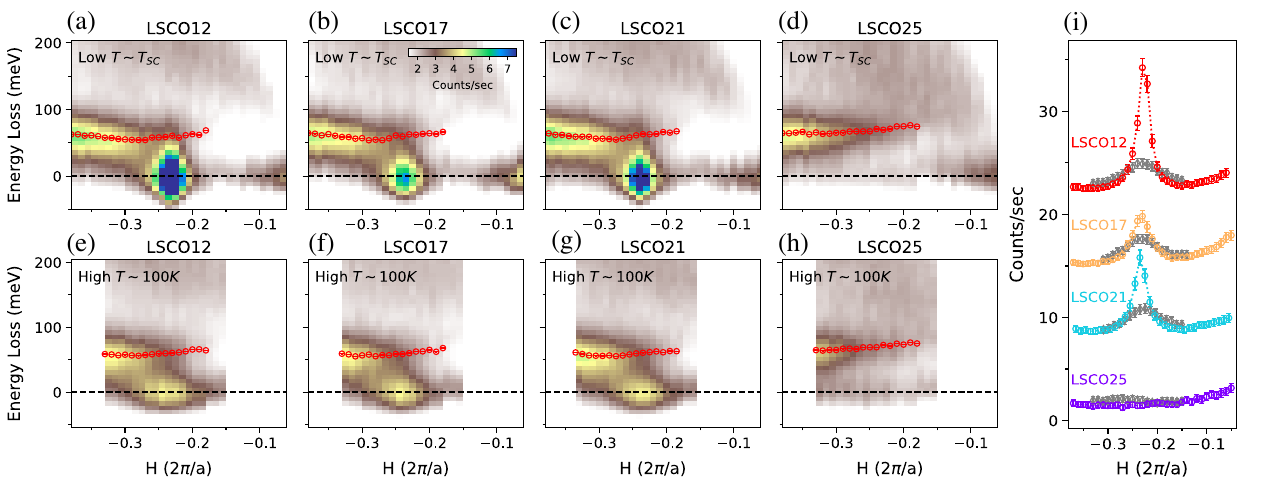}
\caption{Doping and temperature evolution. (a)-(d) show  ($\Delta E=55$~meV) \gls*{RIXS} maps of LSCO$n$ ($n=12$, $17$, $21$, $25$) at their superconducting transition temperature $T_{\text{SC}}$, where the \gls*{CDW} is strongest. All maps share the same colorscale inset in panel (b). Red circles represent the \gls*{BS} phonon energy. (e)-(h) The same measurement for $T = 100$~K. (i) Comparison of integrated intensity in the $\pm20$~meV energy window. Data from panels (a)-(d) at $T = T_\text{SC}$ are represented in color; data at $T = 100$~K from panels (e)-(h) are plotted in gray.}
\label{rixsmapping}
\end{figure*}

In this Letter, we use ultrahigh energy-resolution \gls*{RIXS} to examine the nature of the \gls*{CDW} in \gls*{LSCO}. We observe that \gls*{CDW} correlations and associated \gls*{CDW}-induced phonon-softening persist up to a strikingly high doping level of $x = 0.21$ before both effects disappear at $x = 0.25$. The large doping range $x = 0.12 \rightarrow 0.25$ traverses a topological transition in the Fermi surface \cite{Yoshida2006systematic, Horio2018three, Miao2019discovery}, allowing us to empirically test the relative importance of charge and lattice effects in the excitation spectra. We find that the data can be described entirely in terms of \gls*{CDW}-induced phonon softening and wavevector-dependent changes in the phonon displacement, without invoking any more complex electronic excitations or \gls*{CDW}-related modification of the \gls*{EPC}. Overall, our results suggest a `real-space' picture in which the CDW emerges due to strong electronic correlations and modifies the underlying phonons.

\gls*{LSCO} single crystals with $x=0.12$, $0.17$, $0.21$, and $0.25$ were grown using the floating zone method and are denoted as LSCO$n$ where $n = 12$, $17$, $21$, and $25$, respectively \cite{supp}. Structural and electronic characterization of the samples indicates excellent quality \cite{supp, Miao2019discovery}. High energy-resolution \gls*{RIXS} measurements were performed at 2-ID at the National Synchrotron Light Source II, Brookhaven National Laboratory with a resolution of $\Delta E = 30$~meV \gls*{FWHM} and at I21 at Diamond Light Source featuring a resolution of $\Delta E = 55$~meV. The \gls*{RIXS} process is shown in Fig.~\ref{intro}(a). X-rays were tuned to the Cu $L_3$-edge and measurements were taken with $\sigma$ x-ray polarization perpendicular to the scattering plane (unless otherwise specified). Reciprocal lattice units (r.l.u.) are defined in terms of $(H, K, L)$ with lattice constants $a=b= 3.76$~\AA, $c = 13.28$~\AA. Different values of $H$ and $K$ were accessed by rotating the sample about the $\theta$ and $\chi$ axes without changing the scattering angle $2\theta$. Intensities were normalized to the intensity of the $dd$ excitations similar to previous works \cite{Ghiringhelli2012, Miao2017, Chaix2017, Miao2019a}. Grazing incidence geometry (defined as negative $H$) was chosen to enhance the intensity of charge and lattice (phonon) excitations, and to suppress spin excitations, which are, in any case, $>200$~meV and outside of the energy window we focus on in this paper \cite{DeanLSCO2013, Meyers2017doping}.

Figure ~\ref{intro}(b) plots RIXS data of LSCO at the Cu $L_3$-edge illustrating the main spectral features studied here: (i) a quasi-elastic peak and (ii) a dispersive feature around $50-65$~meV. Feature (i) contains a component of trivial elastically scattered x-rays due to the finite disorder (defects) in the sample and surface scattering. Quasi-elastic scattering is further enhanced by static or quasi-static \gls*{CDW} correlations and displays a peak at \QCDW{} whenever such correlations are present. The inelastic feature (ii) has been seen in several other cuprate \gls*{RIXS} experiments \cite{Chaix2017, Peng2019enhanced, Lucio2019, Matteo2019, Yu2019, Miao2019a} and is assigned to the in-plane Cu-O \acrfull*{BS} phonon mode in agreement with early inelastic x-ray and neutron works \cite{Fukuda2005doping, Reznik2006, Graf2007in, Park2014, Reznik2007, Pintschovius2006Oxygen, Pintschovius1999anomalous}. The spectra were fitted with a Pseudo-Voigt function for the elastic peak, an anti-symmetric Lorentzian function for the \gls*{BS} mode and a linear background. All components of the fit were convoluted with the energy resolution function.

Having assigned the basic spectral features, we use the ultra-high throughput of the I21 beamline to comprehensively map out the momentum and doping dependence of LSCO$n$ $12 \leq n \leq 25$, (see Fig.~\ref{rixsmapping}). This doping range crosses over from $1/8$ doping, where the \gls*{CDW} correlations are strongest, into the overdoped Fermi-liquid-like phase where the \gls*{CDW} disappears \cite{Miao2019discovery, Yoshida2006systematic, Horio2018three}. Importantly, this traverses a topological transition in the electronic structure where hole-pocket/arc-type states transform into an electron-like Fermi surface with a more Fermi-liquid-like scattering rate \cite{Miao2019discovery, Yoshida2006systematic, Horio2018three, Cooper2009anomalous}. This allows us to investigate the relationship of the \gls*{RIXS} spectra with the changes in electronic structure.

We first discuss the quasi-elastic \gls*{CDW} feature, which is summarized in Fig.~\ref{rixsmapping}(i) showing the integrated intensity in the $\pm 20$~meV energy range of Fig.~\ref{rixsmapping}(a)-(h). The intensity enhancement of the resonant process combined with the background suppression attained by energy-resolving the scattered beam make \gls*{RIXS} very sensitive to even very short correlation length \gls*{CDW}s. A \gls*{CDW} around an in-plane wavevector of $(-0.23, 0)$ is not only observed for LSCO12, where it was seen several times previously \cite{Thampy2014, Croft2014, wu2012charge, Wen2019}, but also up to far higher dopings of LSCO21 \footnote{These RIXS results are confirmed by bulk sensitive hard (8~keV) x-ray diffraction measurement under review \cite{Miao2019discovery}}. The $H$-width of the quasi-elastic scattering is consistent with correlation lengths (calculated as 2/FWHM) of $25-45$~\AA{} with shorter values for $x=0.17$. Within error, the peaks exist at the same $H_\text{CDW} = -0.231 \pm 0.005 $ r.l.u., consistent with the stripe phenomenology where  $H_\text{CDW}$ saturates for $x>1/8$ \cite{Yamada1998}. As expected, a substantial fraction of the \gls*{CDW} intensity is suppressed upon warming to 100~K, leaving only a much weaker and diffuse signal \cite{Thampy2014, Croft2014, Miao2019discovery}. We further observe a non-monotonic intensity dependence as function of doping, with LSCO17 being weaker than LSCO12 and LSCO21. There are multiple potential explanations for such a behavior. Perhaps most plausible is to note that the quasi-static \gls*{CDW} intensity in cuprates tends to compete with superconductivity, so the reduced \gls*{CDW} intensity at $x=0.17$ may be associated with the enhanced superconducting correlations that are known to exist at this near-optimal doping level.

Next we discuss the inelastic component of the spectra in Fig.~\ref{rixsmapping}(a-h). In the 100-200~meV energy window, above the maximum phonon energy, flat, structure-less intensity arising from the charge continuum and the tail of the higher energy paramagnon excitations is present over all \Q{}. This intensity shows no clear changes around \QCDW{} and minimal changes with doping. In fact, only a slight increase in the overall intensity is found with increasing doping, which is expected as overdoped samples are more metallic. The inelastic intensity below 100~meV is dominated by the BS phonon, which shows clear energy dispersion and very strong intensity varaiation. Superficial inspection of the raw intensity in Fig.~\ref{rixsmapping}(a) may appear as if a soft mode is dispersing to zero energy at \QCDW{}. This can be examined through a systematic study of samples with different doping levels. To separate the phonon from charge and quasielastic intensity, we fit the data using the previously described model and display the results as red circles in Fig.~\ref{rixsmapping}(a-h). Since the phonon intensity drops strongly as $|H|$ decreases, we focus on $|H| > 0.18$ where the BS phonon can be fitted with good precision.

\begin{figure}
\center
\includegraphics[width = 0.5\textwidth]{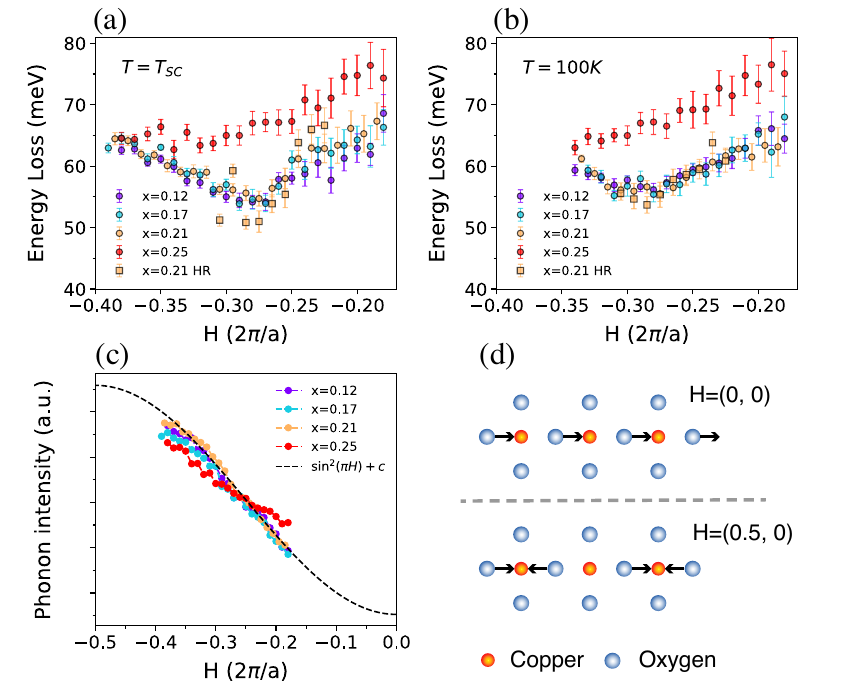}
\caption{Momentum dependent energy and intensity of the \gls*{BS} phonon mode. (a)\&(b) Show the phonon energy dispersion and softening in the vicinity of \QCDW{} at (a) $T=T_\text{SC}$ and (b) $T=100$~K. Circles and squares represent data from Fig.~\ref{rixsmapping} ($\Delta E = 55$~meV) and Fig.~\ref{intro} ($\Delta E = 30$~meV), respectively. (c) Plots the integrated intensity of the \gls*{BS} phonon mode in the [40, 120]~meV energy window after subtracting elastic peak. The dashed line represents the $\sin^2(\pi H) + \text{const.}$ fit to LSCO21. (d) Illustration of the oxygen atom displacements involved in the BS  mode at $H=(0,0)$ and $(0.5,0)$.}
\label{parameters}
\end{figure}

 Figure~\ref{parameters} summarizes the evolution of the BS phonon parameters. Although the phonon softening, shown in panel (a), is appreciable ($13\pm 4$~meV or $19\pm 6\%$), it never shows full soft-mode (i.e.\ zero energy) behavior. The simultaneous disappearance of both the elastic peak and the phonon softening in LSCO25 makes a strong case that the softening is intimately related to the \gls*{CDW} correlations. Inelastic x-ray and neutron scattering measurements of phonon softening in underdoped cuprates have also assigned the phonon
softening to CDW correlations \cite{Pintschovius1999anomalous, Fukuda2005doping, Graf2007in,Reznik2006, Park2014, Reznik2007, Pintschovius2006Oxygen}. It is noted that the phonon energy in LSCO25 is slightly higher than other samples, which might be linked to lattice contraction associated with large Sr concentrations. 

The phonon intensity dispersion is plotted in Fig.~\ref{parameters}(c). Within error, no phonon intensity anomalies are seen around \QCDW{}, instead the clearest feature is a strong increase with $|H|$. Since \gls*{RIXS} excites phonons via the \gls*{EPC} process, the measured intensity reflects this interaction strength and scales with
$g^2$ where $g$ is the \gls*{EPC} \cite{Ament2011, Devereaux2016, Meyers2018decoupling, Lucio2019, Matteo2019}. As a function of \Q{}, the ``breathing-type'' Cu-O bond displacement involved in the BS mode changes. Assuming a well-defined Madelung energy change associated with Cu-O bond stretching, one can predict \gls*{RIXS} intensity scaling  $I \propto g_{\text{br}}^2 =  \sin^{2}(\pi H) + \sin^{2}(\pi K) $ \cite{Johnston2010, Devereaux2016, Lucio2019, Matteo2019}. The comparison in Fig.~\ref{parameters}(c) shows that this simple model is sufficient to describe the intensity behavior of our data, without invoking any more complex phenomenology. It is worth adding that definitively distinguishing $\sin^2(\pi H)$ scaling from other scaling forms is somewhat challenging. The photon energy corresponding to the Cu $L_3$ edge intrinsically limits the highest $|H|$ we can reach, and at low $|H|$ leakage of specular scattering intensity overwhelms the low-energy region of the RIXS spectra. The slight flattening of the dispersion at $x=0.25$ might arise from some leakage of additional background at high doping levels. With these considered, the agreement with $\sin^2(\pi H)$ scaling holds well in the reciprocal-space range measured.

\textit{Discussion of the CDW}.--- Both the quasielastic \gls*{RIXS} intensity and the phonon softening demonstrate the existence of \gls*{CDW} correlations up to a remarkably high doping level of $x=0.21$, traversing the topological transition in the electronic structure \cite{Yoshida2006systematic, Horio2018three, Miao2019discovery}, in which arc or hole-pocket like states centered around the Billouin zone corner transform into an electron-like Fermi surface at the Brillouin zone center. This result confirms very recent non-resonant diffraction measurements and shows, due to the resonant nature of the \gls*{RIXS} probe, that the correlations involve the electronically active Cu states  \cite{Miao2019discovery}. The persistence of the \gls*{CDW} correlations, despite very substantial Fermi surface changes, provides a vivid demonstration that the LSCO \gls*{CDW} cannot be described using any type of weak-coupling Fermi surface nesting picture, as previously suggested for some other cuprate materials \cite{Comin2016}. Instead, the nearly doping-independent \gls*{CDW} wavevector supports mechanisms in which the periodicity of the \gls*{CDW} is set by the short-range electronic interactions. Here, doped holes can save super-exchange energy by clustering together and breaking fewer magnetic bonds, but by doing so, they pay a cost of increased kinetic and Coulomb energy. It has been proposed that a \gls*{CDW} is the optimal compromise between these two tendencies \cite{Emery1990}.  However, a fascinating question remains regarding why the \gls*{CDW} is so stable against increasing doping and electron itineracy. Furthermore, our results motivate a reexamination of the anomalous transport properties of the cuprates, which are often discussed in terms of strange metal physics below a critical doping level of $x_c\approx0.19$, where the systems becomes increasingly more Fermi-liquid-like \cite{Keimer2015}. Since \gls*{CDW} correlations can exist over a more extended doping range than previously thought, it is interesting to consider their influence on transport properties \cite{Seibold2019marginal}. These results, however, argues against a quantum critical point that is associated with a \gls*{CDW} transition as similar as \gls*{CDW} correlations exist in LSCO17 and LSCO21, either side of the putative critical doping of $x_c\approx0.19$. In terms of the \gls*{CDW} fluctuations, we note that the size of the phonon softening is reduced when warming to $T=100$~K, but the magnitude of the reduction is considerably less than the reduction in the quasi-elastic \gls*{CDW} intensity. This suggests that only a relatively small fraction of the total CDW correlations are nucleated into the \gls*{CDW}-order, as otherwise the magnitude of the phonon softening would be expected to scale with the quasi-elastic intensity. 

An intriguing feature of the \gls*{CDW} effects is that the \Q{}-vector with the largest phonon softening does not coincide with the peak in the quasi-elastic scattering at 0.235~r.l.u., but occurs at a larger  wavevector of $H=0.275$~r.l.u. We can consider two candidate explanations for this. The first is that the interactions causing the \gls*{CDW} instability are not the same as the interactions that pin the \gls*{CDW}. Electronic interactions, for instance, may potentially stabilize a wide-range of \gls*{CDW} wavevectors, but lattice- or \gls*{SDW}-coupling might play the final role in setting the final \gls*{CDW} wavevector. This concept was proposed several years ago \cite{Zachar1998} and is further supported by the previously observed change in \gls*{CDW} wavevector with temperature \cite{Miao2017, Miao2018, Miao2019a}. The phonon softening does not show any obvious difference between $T_{SC}$ and $100$~K in Fig.~\ref{parameters}(a)\&(b), in contrast to the \gls*{CDW} peak, which also supports distinct mechanisms for \gls*{CDW}-formation and \gls*{CDW}-pinning. Another potential explanation for the shift is that the effect is due to the $\Q$ dependence of the \gls*{EPC}. Since \gls*{EPC} increases with $H$, the softening at higher $H$ will be enhanced, which would displace the point of maximum softening.

\textit{Discussion of electronic CDW excitations}. --- A strength of the current dataset is the opportunity to compare LSCO12 and LSCO21, which exhibit comparable levels of \gls*{CDW} order, despite their very different doping levels. Any electronic \gls*{CDW} excitations that may be present in the RIXS spectra would be expected to increase with electronic density of states and change substantially with doping. Since the overall form of the spectra is rather similar over this large doping range, our results argue against the presence of electronic \gls*{CDW} excitations, as suggested previously, as a ubiquitous, intrinsic feature of \gls*{RIXS} spectra of the cuprates \cite{Chaix2017}. It is, however, possible that the x-ray polarization is important to observing electronic \gls*{CDW} excitations \footnote{Data in Ref.~\cite{Chaix2017} is taken with positive $H$ (grazing exit geometry), district from the negative $H$ (grazing incidence geometry).}.

\textit{Electron phonon coupling}. --- The concept of using \gls*{RIXS} to extract \gls*{EPC} in cuprates has generated considerable excitement recently \cite{ament2011determining, Devereaux2016, Matteo2019, Lucio2019, Peng2019enhanced}. Our results support this, in the sense that we observe intensity scaling as $I\propto \sin^{2} (\pi H)$. However, no phonon intensity anomalies related to the \gls*{CDW} are observed. A recent preprint reports measurements of  La$_{1.8-x}$Eu$_{0.2}$Sr$_x$CuO$_{4+\delta}$ which used an incident energy detuning method, formulated in Refs.~\cite{ament2011determining, Matteo2019, Lucio2019}, to suggest a very large \gls*{CDW}-induced modification of the \gls*{EPC} from $g=0.30 \rightarrow 0.35$ upon cooling into the \gls*{CDW} phase \cite{Peng2019enhanced}. Assuming $I\propto g^2$, this would imply a phonon intensity change of $(0.35/0.3)^2 \sim 1.4$. As such, it is very difficult to justify the absence of a discernible $\sim40$\% temperate-induced change in the phonon dispersion at \QCDW{} on resonance in our data or that in Ref.~\cite{Peng2019enhanced}.

\textit{RIXS as a probe of cuprate CDWs}. --- Overall, our results support a real-space picture of an electronically driven \gls*{CDW}, without needing to invoke nesting or a van Hole singularity. RIXS measures the phonon softening occuring due to the charge modulation, but the overall doping dependence is fully explicable without invoking more complex phason, Fano, or \gls*{CDW}-enhanced \gls*{EPC} effects that have generated considerable excitement recently \cite{Dean2013, Miao2017, Chaix2017, Arpaia2019, Peng2019enhanced}. Our observation is backed, by improved energy resolution ($\Delta E = 30$~meV) and more extensive doping dependence compared with what has been done previously \cite{Dean2013, Miao2017, Chaix2017, Arpaia2019, Peng2019enhanced}. Excluding these exciting effects is at some level disappointing in terms of novel excitations, but is, however, highly important in view of the more-and-more extensive use of \gls*{RIXS}. Although \gls*{RIXS}, in this case, provides information similar to inelastic x-ray and neutron scattering, compelling applications for \gls*{RIXS} remain in cases where the x-ray penetration depth and resonant mode selectivity is important \cite{Meyers2018decoupling}.

In conclusion, we report \gls*{RIXS} measurements of \gls*{CDW} correlations in \gls*{LSCO} over an extensive doping range. \gls*{CDW}-related quasi-elastic scattering and phonon softening is observed from $x=0.12$ to $x=0.21$, traversing a topological transition in the Fermi-surface, before disappearing at $x=0.25$. Based on these results, we conclude that the spectra have little or no direct coupling to electronic excitations. Instead the spectra are dominated by \gls*{CDW}-driven phonon softening and phonon intensity variations arising from changes in the phonon displacement as a function of \Q{}. Overall, our results support a senario in which the \gls*{CDW} is driven by strong correlations and clarify that the low-energy \gls*{RIXS} response in cuprates are driven by the \gls*{CDW} modifying the lattice, invoking more complex interactions. 

\begin{acknowledgments}
This material is based upon work supported by the U.S. Department of Energy, Office of Basic Energy Sciences, Early Career Award Program under Award No. 1047478. Work at Brookhaven National Laboratory was supported by the U.S.\ Department of Energy, Office of Science, Office of Basic Energy Sciences, under Contract No.~DE-SC0012704. X. L.\ and J.Q.L.\ were supported by the ShanghaiTech University startup fund, MOST of China under Grant No. 2016YFA0401000, NSFC under Grant No.\ 11934017 and the Chinese Academy of Sciences under Grant No.\ 112111KYSB20170059. This research used resources at the Soft Inelastic X-Ray beamline of the National Synchrotron Light Source II, a U.S. Department of Energy (DOE) Office of Science User Facility operated for the DOE Office of Science by Brookhaven National Laboratory under Contract No.~DE-SC0012704. We acknowledge Diamond Light Source for time on Beamline I21 under Proposal 22261.
\end{acknowledgments}

\bibliography{ref}
\end{document}


\title{Supplemental Material: Nature of the charge density wave excitations in cuprates} 

\author{J. Q. lin}\email[]{jiaqilin@bnl.gov}
\affiliation{Condensed Matter Physics and Materials Science Department, Brookhaven National Laboratory, Upton, New York 11973, USA}
\affiliation{School of Physical Science and Technology, ShanghaiTech University, Shanghai 201210, China}
\affiliation{Institute of Physics, Chinese Academy of Sciences, Beijing 100190, China
}
\affiliation{University of Chinese Academy of Sciences, Beijing 100049, China}
\author{H. Miao}
\altaffiliation[Present address: ]{Materials Science and Technology Division, Oak Ridge National Laboratory, Oak Ridge, Tennessee 37831, USA}
\affiliation{Condensed Matter Physics and Materials Science Department, Brookhaven National Laboratory, Upton, New York 11973, USA}
\author{D. G. Mazzone}
\author{G. D. Gu}
\affiliation{Condensed Matter Physics and Materials Science Department, Brookhaven National Laboratory, Upton, New York 11973, USA}

\author{A. Nag}
\author{A. C. Walters}
\author{M. Garc\'{i}a-Fern\'{a}ndez}
\affiliation{Diamond Light Source, Harwell Science and Innovation Campus, Didcot, Oxfordshire OX11 0DE, United Kingdom}

\author{A. Barbour}
\author{J. Pelliciari}
\author{I. Jarrige}
\affiliation{National Synchrotron Light Source II, Brookhaven National Laboratory, Upton, NY 11973, USA}

\author{M. Oda}
\author{K. Kurosawa}
\affiliation{Department of Physics, Hokkaido University, Sapporo 060-0810, Japan}
\author{N. Momono}
\affiliation{Department of Sciences and Informatics, Muroran Institute of Technology, Muroran 050-8585, Japan}

\author{K. Zhou}
\affiliation{Diamond Light Source, Harwell Science and Innovation Campus, Didcot, Oxfordshire OX11 0DE, United Kingdom}
\author{V. Bisogni}
\affiliation{National Synchrotron Light Source II, Brookhaven National Laboratory, Upton, NY 11973, USA}
\author{X. Liu}\email[]{liuxr@shanghaitech.edu.cn}
\affiliation{School of Physical Science and Technology, ShanghaiTech University, Shanghai 201210, China}
\author{M. P. M. Dean}\email[]{mdean@bnl.gov}
\affiliation{Condensed Matter Physics and Materials Science Department, Brookhaven National Laboratory, Upton, New York 11973, USA}

\date{\today}

\maketitle

\renewcommand{\thefigure}{S\arabic{figure}}

This document provides additional details of the sample synthesis, sample characterization, temperature-dependent phonon intensity data, and fitting of the RIXS spectra.

\section{Sample synthesis}
La$_{2-x}$Sr$_x$CuO$_4$ samples were grown by the traveling-solvent floating-zone method. For each composition, a single feed rod with a length of 20--25~cm was used. After the growth, the first few centimeters of the crystal rod were removed and discarded, while the remainder was annealed in flowing O$_{2}$ at $980^{\circ}$C for 1 week. Sample surfaces were prepared by cleaving the samples mechanically to expose a $c$-axis face. The samples were crystallographically aligned prior to the measurements RIXS using Laue diffraction. During the RIXS measurements themselves, the $\theta$ and $\chi$ angles were further refined by optimizing the CDW intensity as a function of both angles.

 \section{Sample characterization}
 The superconducting transition temperatures were determined by magnetization measurements in a 1~mT applied magnetic field (after cooling in zero field) yielding the expected values of $T_{\text{SC}}=$ 28, 37, 30, and 10~K, for LSCO12, LSCO17, LSCO21 and LSCO25, respectively. X-Ray diffraction measurements confirmed excellent sample crystallinity with crystal mosaics of the order of $0.02^{\circ}$. The doping level was characterized by angle-resolved photoemssion spectroscopy (ARPES)  \cite{Miao2019discovery}. The effective doping was determined to be consistent with the Sr concentration via a tight binding fitting of the Fermi surface. The  topological change in Fermi surface topology was also observed between LSCO17 and LSCO21. Very similar results have been observed in previous ARPES studies of LSCO \cite{Yoshida2006systematic, Horio2018three}.

\section{Phonon intensity dependence}
\begin{figure*}
\includegraphics[width = .7\textwidth]{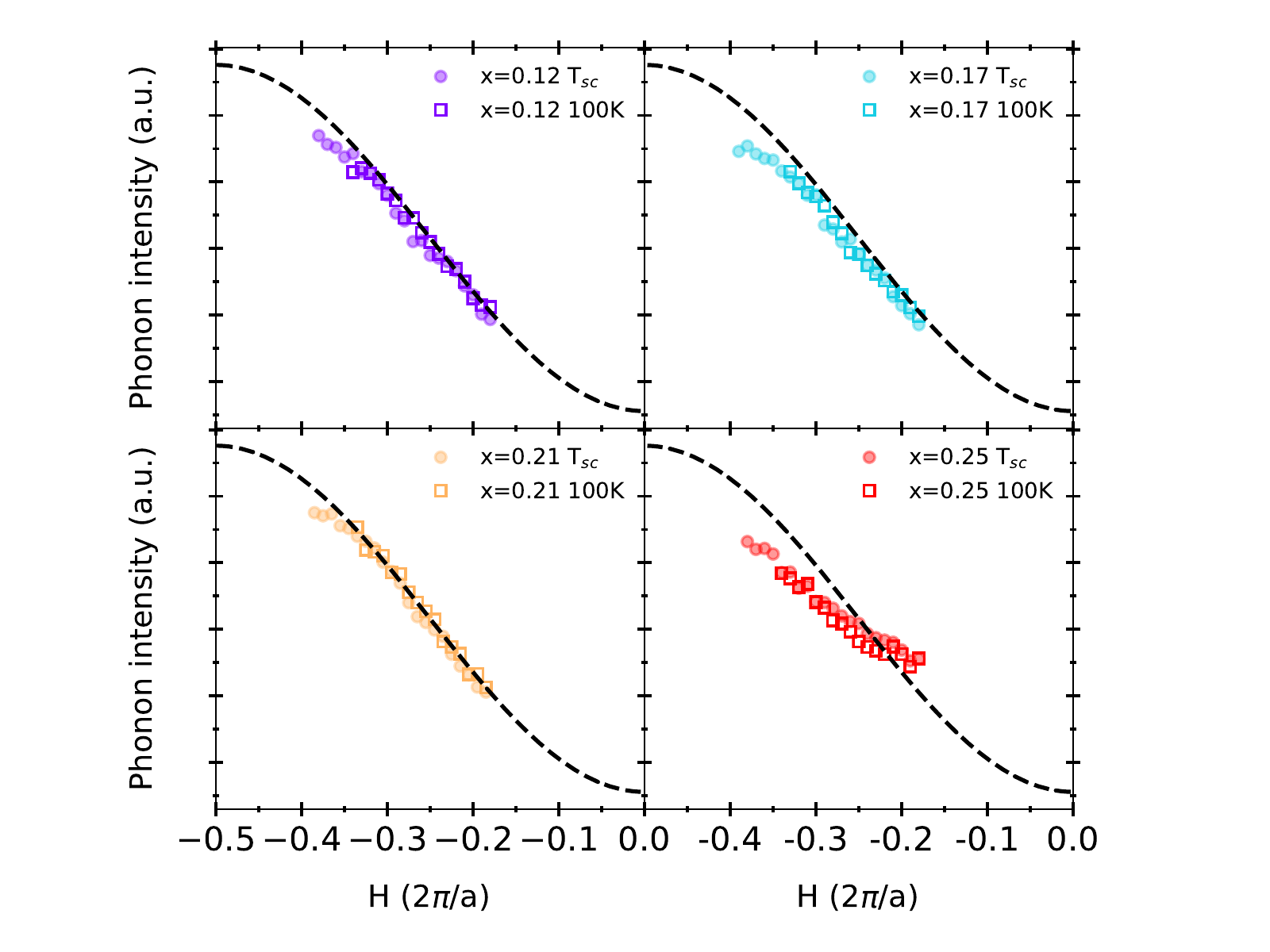}
\caption{Phonon intensity as a function of $H$ similar to Fig.~3(c) of the main text, but with the 100~K data overlaid on the low-temperature data and separated onto different panels. This shows that the phonon intensity is the same at $T_c$ and 100~K for all the different dopings.}
\label{Tdep_phonon_intensity}
\end{figure*}

In Fig.~\ref{Tdep_phonon_intensity} we show that the phonon intensity dispersion is essentially identical at $T_c$ and 100~K.

\section{Fitting of the RIXS spectra}
\begin{figure*}
\includegraphics[width = .95\textwidth]{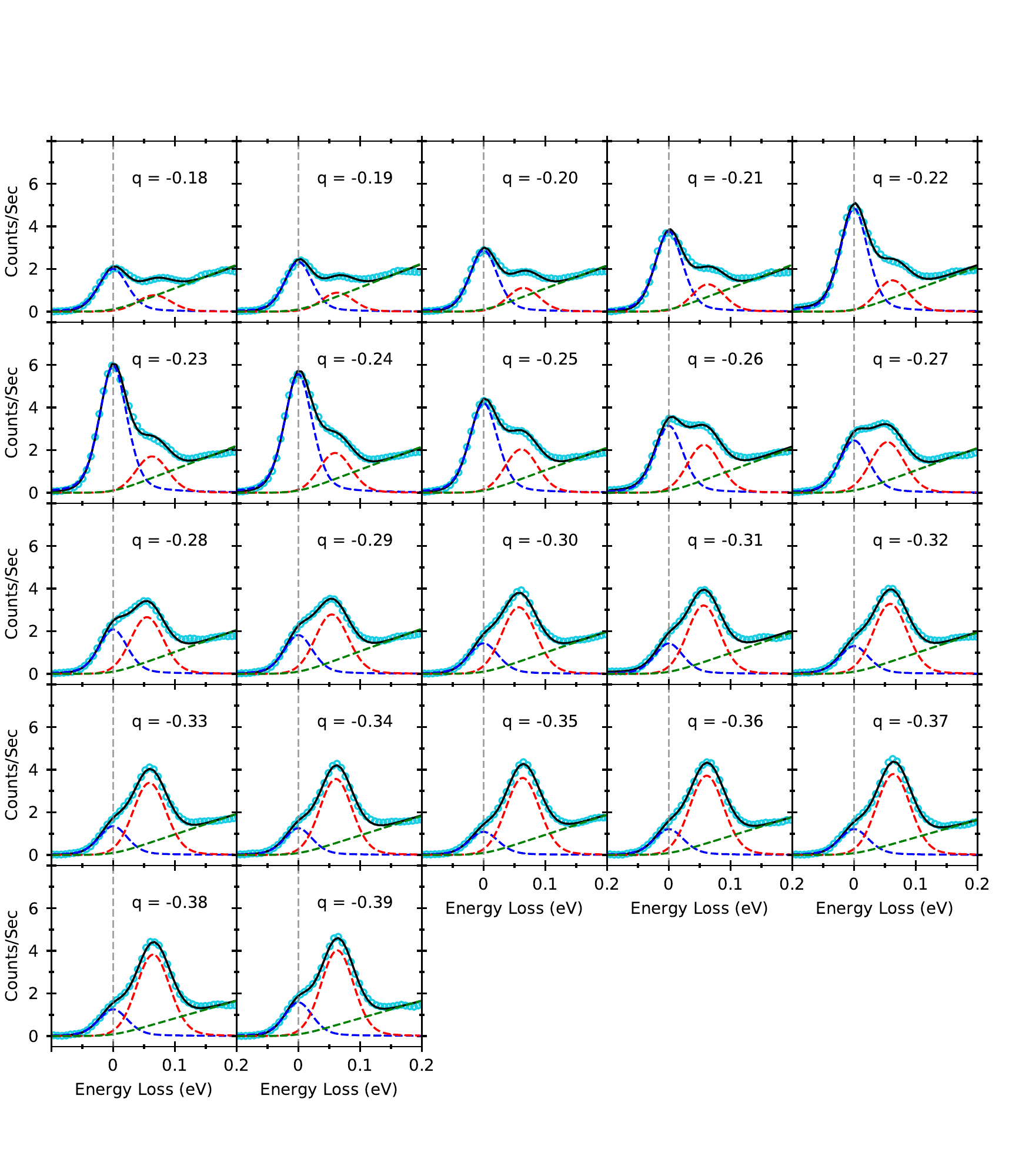}
\caption{Fitting results for LSCO17 at $T=T_{SC}$. The data is represented by cyan dots and the fit is shown as a black line. Blue, red and green dashed lines represent the elastic line, phonon mode and background components, respectively.}
\label{show_fitting_p17_lowT}
\end{figure*}

In order to illustrate the method and quality of data fitting, we
 present the fitting result for LSCO17 at low temperature in Fig.~\ref{show_fitting_p17_lowT}.  The spectra were fitted with a Pseudo-Voigt function for the elastic peak, an anti-symmetric Lorentzian function for the \gls*{BS} mode and a linear background, all convoluted with resolution.

\bibliography{ref}